\pdfoutput=1
\documentclass[11pt]{article}
\usepackage[margin=1in]{geometry}
\usepackage{setspace} 
\usepackage{layouts}
\usepackage{textcomp}

\usepackage{algorithm}
\usepackage{algorithmic}
\usepackage{multirow}
\usepackage[shortcuts]{extdash} 
\usepackage{tikz}
\usepackage{xcolor}
\usepackage{booktabs}
\usepackage{tabularx}
\usepackage{graphicx}
\usepackage{nicefrac}
\usepackage{hhline}
\usepackage{makecell}
\usepackage{bm}
\usepackage{bbm}
\usepackage{mathtools}
\usepackage{amsmath}
\usepackage{amsthm}
\usepackage{amssymb}
\usepackage{float}
\usepackage{subfigure}
\usepackage{mwe}
\usepackage{makecell}
\usepackage{appendix}
\usepackage[section]{placeins}
\usepackage{natbib}
\setcitestyle{authoryear}

\theoremstyle{plain}
\usepackage{authblk}

\DeclarePairedDelimiter{\abs}{|}{|}
\DeclarePairedDelimiter{\brac}{[}{]}

\DeclarePairedDelimiter{\paren}{\lparen}{\rparen}

\newcommand{\mc}[1]{\ensuremath \mathcal{#1}}
\newcommand{\mb}[1]{\ensuremath \mathbb{#1}}

\newcommand{\old}[1]{\ensuremath \bar{#1}}
\newcommand{\new}[1]{\ensuremath \ddot{#1}}

\newcommand{\SN}{\textup{SN}}

\newcommand{\TN}{\textup{TN}}
\newcommand{\normal}{\textup{N}}

\newcommand{\smpOne}{\mb{S}_1}
\newcommand{\smpTwo}{\mb{S}_2}

\newcommand{\real}{\mb{R}}

\newcommand{\EE}{\mathbb{E}}

\newcommand{\ccdot}{*}

\newcommand{\rowname}[1]
{\rotatebox{90}{\makebox[\tempheight][c]{\emph{#1}}}}

\newcommand{\ylabel}[1]
{\rotatebox{90}{\makebox[\tempheight][c]{\tiny{#1}}}}

\newcommand{\columnname}[1]
{\makebox[\tempwidth][c]{#1}}

\title{\textbf{New mixture models for decoy-free false discovery rate estimation in mass-spectrometry proteomics}}

\author{Yisu Peng\,$^{1,\dagger}$, Shantanu Jain\,$^{1,\dagger}$, Yong Fuga Li\,$^{2}$, Michal Gregu{\v{s}}\,$^{3,4}$, Alexander R. Ivanov\,$^{3, 4}$, Olga Vitek\,$^{1,4}$ and Predrag Radivojac\,$^{1,*}$}

\date{} 

\begin{document}

\maketitle
\vspace{-2 em}
\noindent {\footnotesize $^{\text{\sf 1}}$Khoury College of Computer Sciences, Northeastern University, Boston, Massachusetts, U.S.A.; $^{\text{\sf 2}}$Illumina Inc., San Diego, California, U.S.A.; $^{\text{\sf 3}}$Department of Chemistry and Chemical Biology, Northeastern University, Boston, Massachusetts, U.S.A.; $^{\text{\sf 4}}$Barnett Institute of Chemical and Biological Analysis, Northeastern University, Boston, Massachusetts, U.S.A.}\\
\noindent{\footnotesize $^\dagger$Contributed equally to this work.; $^\ast$To whom correspondence should be addressed.}
\vspace{1 em}

\begin{abstract}
\noindent
\textbf{Motivation:} Accurate estimation of false discovery rate (FDR) of spectral identification is a central problem in mass spectrometry-based proteomics. Over the past two decades, target-decoy approaches (TDAs) and decoy-free approaches (DFAs), have been widely used to estimate FDR. TDAs use a database of decoy species to faithfully model score distributions of incorrect peptide-spectrum matches (PSMs). DFAs, on the other hand, fit two-component mixture models to learn the parameters of correct and incorrect PSM score distributions. While conceptually straightforward, both approaches lead to problems in practice, particularly in experiments that push instrumentation to the limit and generate low fragmentation-efficiency and low signal-to-noise-ratio spectra. \\
\textbf{Results:} We introduce a new decoy-free framework for FDR estimation that generalizes present DFAs while exploiting more search data in a manner similar to TDAs. Our approach relies on multi-component mixtures, in which score distributions corresponding to the correct PSMs, best incorrect PSMs, and second-best incorrect PSMs are modeled by the skew normal family. We derive EM algorithms to estimate parameters of these distributions from the scores of best and second-best PSMs associated with each experimental spectrum. We evaluate our models on multiple proteomics datasets and a HeLa cell digest case study consisting of more than a million spectra in total. We provide evidence of improved performance over existing DFAs and improved stability and speed over TDAs without any performance degradation. We propose that the new strategy has the potential to extend beyond peptide identification and reduce the need for TDA on all analytical platforms.\\
\textbf{Availability:} {https://github.com/shawn-peng/FDR-estimation} \\
\end{abstract}

\section{Introduction}
A typical bottom-up proteomics pipeline consists of several experimental and computational steps, combined to interrogate the presence, quantity, form, and function of proteins in the biological mixture \citep{Aebersold2003, Steen2004, Gingras2007, Choudhary2010}. Central to all these challenges is the task of accurately establishing the presence of peptide species in the  sample \citep{Kall2008b, Hubler2020}, a step that relies on computational and statistical techniques to map spectra from the mass spectrometer to peptide sequences and assign confidence scores to the resulting peptide-spectrum matches (PSMs). Peptide identification is often performed via a search algorithm, where experimental spectra are scored against the theoretical spectra derived from a selected group of candidate peptides \citep{Yates1995, Perkins1999, Tabb2007, Kim2014, Kong2017} or \emph{de novo}, when restricting the set of candidate peptides is problematic \citep{Dancik1999, Frank2005}.

Despite methodological variability in practice, the core of any peptide identification protocol is the scoring of PSMs that is intended to reflect their likelihood of being correct assignments \citep{Li2012, Hubler2020}. These schemes must meet both local and global requirements in that the ranking of PSMs for a given experimental spectrum must prioritize the most likely peptide assignments and that the scoring of those top-ranked PSMs over all experimental spectra must be calibrated so that the global ranking of top-ranked PSMs is meaningful \citep{Keich2015}. Well-performing search engines generally meet these requirements, in which case the set of identified or accepted PSMs can be reliably determined from the ranked list of top-scoring PSMs based on a score threshold. The list of identified PSMs ideally contains a large fraction of correct identifications (spectra matched to peptides they originated from) and not more than a small fraction of incorrect identifications (spectra matched to peptides they did not originate from).

False discovery rate (FDR) is defined as the expected proportion of incorrect identifications among reported identifications \citep{Storey2002, Choi2008, Burger2018}. Over the past two decades, two major approaches for estimating FDR have emerged; i.e., target-decoy approaches (TDAs) and decoy-free approaches (DFAs). Target-decoy techniques search both the set of peptides possibly present in the sample (target database) and a set of peptides that are not in the sample (decoy database), where the role of the decoy database is to faithfully model the score distribution of incorrect top-scoring PSMs from the target database and thus facilitate FDR estimation \citep{Elias2007}. TDAs differ in the construction of decoy sequences and search strategies such as separately or combined with target sequences \citep{Jeong2012}. Decoy-free techniques, on the other hand, search only the target database and fit a generative two-component model to the set of scores corresponding to all top-scoring PSMs. The two components model the correct and incorrect score distributions, typically using some combination of Gaussian, Gumbel, and Gamma distributions. For example, \cite{Keller2002} model the score distribution of the correct top PSMs using a Gaussian distribution and incorrect top PSMs using a Gamma distribution. An expectation-maximization (EM) algorithm is applied to estimate the parameters of these distributions \citep{Dempster1977}.

Each search strategy comes with pros and cons. Owing to its simplicity, TDA with a concatenated database search has dominated bottom-up proteomics, even if the benefits of competing decoy peptides with target peptides for experimental spectra are incompletely understood. In fact, the usefulness of TDA has been continuously challenged on several grounds \citep{Kim2008, Kall2008, Gupta2011, Cooper2011, Cooper2012, Danilova2019}, including the construction of decoy sequences, choice of FDR estimators, and run time. Current practices generally rely on peptide reversal within each protein to construct decoys, based on empirical characterizations against the alternatives \citep{Elias2007}. TDAs estimate FDR as the fraction of the number of decoy top PSMs and the number of target top PSMs above the threshold. While this approach is reasonable with large datasets, it is theoretically problematic as it can lead to FDR estimates above 1 and possibly even infinity. TDAs also consider protein databases twice in size, which can be computationally expensive for identifying post-translationally modified peptides or cross-linked peptides \citep{Rinner2008, Ji2016}. On the other hand, DFAs are not without problems either. While theoretically pleasing, these methods suffer from restrictive modeling assumptions as well as difficulties in resolving overlapping score distributions, especially when the fraction of correct PSMs is small \citep{Ma2012}. They also lead to inconsistencies, such as ones where Gaussian-Gamma distributions give best fits on average yet the component densities have different supports and can lead to pathological situations; e.g., low-scoring PSMs might have a probability of 1 to be correct \citep{Li2008}. This is particularly problematic in experiments where distinguishing correct and incorrect PSMs is challenging.

The objective of this study is to introduce and explore new decoy-free FDR estimation procedures that combine the strengths of TDAs and DFAs. Specifically, we consider a two-sample approach, where the top or best-scoring PSMs are used in a manner similar to conventional DFA searches, and the second-best PSMs, much like decoy PSMs, are used to improve modeling of the incorrect top PSMs. We model the set of component densities using a relatively new family of skew normal distributions that offer desirable flexibility within the unimodal family yet provide elegant update rules for an EM-based optimization. We evaluate the new systems against both TDAs and DFAs on NIST spectral libraries from four species, ten additional PRIDE datasets from six species as well as an in-house case study using nanogram levels of total HeLa cell digest to demonstrate the potential for applications in high-sensitivity proteomics profiling. We demonstrate that leveraging the extra search information increases the accuracy and the stability of estimates, in particular in experiments where low amounts of biological material limit the quality and the number of spectra \citep{Li2015, Budnik2018}. Overall, we believe that the new algorithms have a potential to generalize beyond peptide identification to all types of search problems involving analytical platforms.

\section{Background}
\subsection{Terminology and notation}
Let $\mathcal{X}=\left\{ x_{i}\right\}$ be a set of spectra collected from a mass spectrometer and $\mathcal{P}=\left\{ p_{j}\right\}$ a set of candidate peptides that are possibly present in the biological sample. A search engine produces a set of triplets $(x, p, s) \in \mathcal{X} \times \mathcal{P} \times \mathbb{R}$, where $s$ is the score assigned to the PSM $(x,p)$. The higher the score, the more likely that the spectrum $x$ was generated from $p$. 

Let now $x$ be generated from some (unknown) peptide $q$ and let $\left((x,p_{1}, s_{1}), (x,p_{2}, s_{2}), \ldots\right)$ be a ranked list of PSMs from a search engine for $x$ such that $s_{1} \geq s_{2} \geq \ldots$ A PSM $(x,p)$ for which $p=q$ is called the \emph{correct match}, whereas all other PSMs involving $x$ are called \emph{incorrect matches}. Furthermore, given the list $\left((x,p_{1}, s_{1}), (x,p_{2}, s_{2}), \ldots\right)$, the PSM with the highest score, $(x,p_{1})$, is called the top, first or best-scoring PSM, the second-ranked PSM, $(x,p_{2})$, is called the second PSM, etc. Finally, we also distinguish among incorrect PSMs. The highest-scoring incorrect PSM for $x$ will be referred to as the top, first or best incorrect PSM, whereas the second-best incorrect PSM will be referred to as the second incorrect PSM.

To reduce complexity, an MS/MS analysis pipeline often keeps only top PSMs for the set of spectra $\mathcal{X}$; i.e., only the top-scoring PSM for each spectrum $x$. It then determines a threshold $\tau$ such that the peptide $p$ from each top hit $(x,p)$ is considered \emph{identified} when the score $s$ from $(x,p,s)$ satisfies $s \geq \tau$. If, further, $p=q$, $p$ is considered to be the correct identification. The threshold $\tau$ can be set based on experience with particular search engines although the most rigorous approach is to estimate FDR for the set of identified peptides obtained by thresholding at $\tau$. Current approaches restrict the analysis to top-scoring PSMs for each experimental spectrum. In this study, we remove this restriction and include both top PSMs and second-best PSMs to more confidently model the data distributions.

\subsection{Skew normal family}

The Gaussian family is widely used in many applications to model real-world data. However, the symmetry of the Gaussian density makes it an inferior choice for modeling skewed data. One approach to account for the skewness is to use a mixture of Gaussian distributions; however, finite Gaussian mixtures are ill-equipped to model the skewness, especially when the data is expected to be unimodal \citep{Jain2019}. In such cases one may choose from one of the many skewed families such as Gumbel, Gamma, Weibull and skew normal. The use of Gumbel and Gamma distributions in the context of FDR estimation has been extensively studied \citep{Li2008}. In this paper, we explore the appropriateness of the skew normal family for FDR estimation. Skew normal family is an appealing choice for modeling competition since the density of the maximum of two identically distributed Gaussian random variables is exactly skew normal \citep{Arellano2006}. 

The univariate skew normal (SN) family was introduced as a generalization of the normal family \citep{azzalini1985class}. It has a location ($\mu$), a scale ($\omega$), and a shape ($\lambda$) parameter, where $\lambda$ controls the direction and degree of skewness. The distribution is right-skewed when $\lambda>0$, left-skewed when $\lambda<0$, and reduces to a normal distribution when $\lambda=0$. The probability density function (pdf) of a random variable $X \sim \SN(\mu,\omega,\lambda)$ is given by
$$f_{\SN}(x;\mu,\omega,\lambda)=\frac{2}{\omega}\phi\paren*{\frac{x-\mu}{\omega}}\Phi\paren*{\frac{\lambda(x-\mu)}{\omega}},\  x \in \real,$$
\noindent where $\mu,\lambda \in \real$, $\omega \in \real^+$, $\phi$ and $\Phi$ are the probability density function (pdf) and the cumulative distribution function (cdf) of the standard normal distribution $\normal(0,1)$, respectively. The cumulative distribution function of $X$ is given by 
$$F_{\SN}(x;\mu,\omega,\lambda)=\Phi\paren*{\frac{x-\mu}{\omega}} -2\mc{T}\paren*{\frac{x-\mu}{\omega}, \lambda},\  x \in \real,$$
where $\mc{T}(h,a)$ is Owen's T function \citep{young1974algorithm}. The \SN\ family can be alternatively parameterized by $\Delta$ and $\Gamma$ instead of $\lambda$ and $\omega$, as defined in Table \ref{tab:altPar}. The alternate parametrization naturally arises in the stochastic representation of a \SN\ random variable: 
\begin{equation}
\label{eq:probSN}
X\sim \SN(\mu,\omega,\lambda) \,\,\Rightarrow \,\,X \overset{d}{=} \mu + \Delta T + \Gamma^{\nicefrac{1}{2}}U,    
\end{equation} 
\noindent where $T\sim \TN(0,1,\real_{+})$, the standard normal distribution truncated below $0$; $U \sim \normal(0,1)$, the standard normal distribution; and $\overset{d}{=}$ reads as ``equal in distribution''. The stochastic representation is useful for deriving many properties of the skew normal distribution and is also used in an EM-based maximum-likelihood estimation \citep{lin2007finite}. The algorithms for the skew normal mixture models derived in this paper also exploit this stochastic representation.

\begin{table}
\caption{\small Alternate parametrization for the skew normal distribution. Update equations of the algorithm are better formulated in terms of the alternate parameters. The table gives the relationship between the alternate and the canonical parameters as well as additional related quantities.}
\vspace{2mm}
\begin{center}
\begin{tabularx}{0.75\textwidth}{|c|c|X|}
    \cline{1-3}
     \multicolumn{2}{|>{\hsize=2\hsize}c|}{ \textbf{Alternate Parametrization}} & \multirow{2}{*}{\textbf{Related Quantities}} \\\cline{1-2}
       canonical $\rightarrow$ alternate & alternate $\rightarrow$ canonical &  \\ 
       \hhline{===}
       \makecell{ $\Delta=\omega\delta$\\ $\Gamma= \omega^2 -\Delta^2 $} &\makecell{ $\lambda = \textrm{sign}(\Delta)\sqrt{\nicefrac{\Delta^2}{\Gamma}}$\\$\omega = \sqrt{\Gamma + \Delta^2 }$\\} & $\delta = \frac{\lambda}{\sqrt{1+\lambda^2}}$\\
       \cline{1-3}
\end{tabularx}
\end{center}
\label{tab:altPar}
\end{table}

\section{Methods}
\label{sec:methods}
In this section we introduce two generative models and derive corresponding EM algorithms for parameter estimation. Let $\smpOne$ denote the set of the first scores and $\smpTwo$ denote the set of the second scores of a tandem mass spectrometry (MS/MS) search. The first model relies solely on the score distributions of the top PSMs and thus only $\smpOne$ is used for parameter estimation. The second model is an extension when first and second PSMs are both considered and uses $\smpOne$ and $\smpTwo$ to estimate the parameters. The dataset sizes $|\smpOne|$ and $|\smpTwo|$ need not be equal.

We assume in both models that the scores corresponding to a correct match and all incorrect matches follow skew normal distributions. Technically, we introduce $C$, $I_1$ and $I_2$ to denote the random variables corresponding to the scores of the correct match, the first incorrect match and the second incorrect match, respectively, as
\begin{equation}
    C \sim \SN\paren*{\theta_c} \,\,\,\,\,\,\,\,\,\,\,\,
    I_1 \sim \SN\paren*{\theta_1}, \,\,\,\,\,\,\,\,\,\,\,\, 
    I_2 \sim \SN\paren*{\theta_2}, 
    \label{mod:corIncor}
\end{equation}
\noindent where $\theta$ denotes the skew normal parameters $\mu$, $\omega$, and $\lambda$.

Sections \ref{sec:1s2dmix}-\ref{sec:2s3dmix} below present only update rules of the proposed EM algorithms. We direct the reader to Supplementary Materials for additional details. Specifically, Section S2 of the Supplementary Materials shows the derivation of the algorithms and Section S1 gives proofs of the supporting lemmas.

\begin{table}
\caption{\small Useful quantities. The parameter update equations are given in terms quantities defined below. The quantities accented with $\old{\ }$ have $\old{\zeta}$, the current estimate of the model parameters, as an implicit parameter. $\old{\zeta}$ contains all the model parameters: $\alpha$ and/or $\beta$ and the parameters for the skew normal components, $\theta_{\ccdot}$; depending upon the model, $\ccdot$ can take values $c,1$ and $2$. $\theta$ contains skew normal parameters $\mu,\omega$ and $\lambda$. Parameters $\delta,\Delta$ and  $\Gamma$ are related to $\omega$ and $\lambda$ as per Table \ref{tab:altPar}. \TN$(\mu,\sigma^2,\real^+)$ represents truncated normal distribution truncated below $0$. $\EE$ represents the expectation operator. The expectations of the first two moments of the \TN\ random variable can be computed as shown in Lemma 1 in the Supplementary Materials.
}
\begin{center}
\begin{tabularx}{0.6\textwidth}{|X|}
\cline{1-1}
\textbf{Quantities} \\\hhline{=} 
\makecell{$\begin{aligned}
     \old{m}_{\ccdot}(x,\Delta)&= x-v(x,\old{\theta}_{\ccdot})\Delta\\
            \old{d}_{\ccdot}(x,\mu)&= v(x,\old{\theta}_{\ccdot})(x-\mu)\\
            \old{g}_{\ccdot}(x,\mu,\Delta)&= (x-\mu)^2 -2\Delta v(x,\old{\theta}_{\ccdot})(x-\mu) + \Delta^2 w(x,\old{\theta}_{\ccdot})\\
            v(x,\theta)&= \EE\brac*{T_x}\\
            w(x,\theta)&= \EE\brac*{T_x^2}\\
            T_x &\sim  \TN\paren*{\nicefrac{\delta}{\omega}(x-\mu), 1-\delta^2,\real^+}
\end{aligned}$}
        \\ \cline{1-1}
\end{tabularx}
\end{center}
\label{tab:algoQuantities}
\end{table}

\subsection{Top score skew normal mixture}
\label{sec:1s2dmix}
The top-score skew normal mixture, referred to as 1SMix model, is the conventional decoy-free model in which both component distributions are in the skew normal family. More formally, we model the first score $S_1$ as a mixture of the correct and first incorrect scores, each being a skew normal random variable; i.e.,
    $$S_1 \sim \alpha \SN(\theta_c) + (1-\alpha) \SN(\theta_1).$$
The triple $\zeta=(\alpha, \theta_c, \theta_1)$ gives the parameters of the model. We obtain the maximum likelihood estimates of $\zeta$ from $\smpOne$ using the EM algorithm for finite skew normal mixture estimation in \cite{lin2007finite}. For completeness, we give a derivation of the algorithm for the two component mixture case in Section S\ref{sec:methods}. Using $\new{\ }$ and $\old{\ }$ to accent the new and old parameters, respectively, the parameter update equations of the EM algorithm are as follows:
\begin{align*}
\new{\alpha} &= \frac{1}{\abs*{\smpOne}}\sum_{s_1\in\smpOne} \old{p}_1(s_1)\\
\new{\mu}_c &= \frac{\sum_{s_1\in\smpOne}\old{p}_c(s_1)\old{m}_c\paren*{s_1,\old{\Delta}_c}}{\sum_{s_1\in\smpOne}\old{p}_c(s_1)}\\
\new{\mu}_1 &= \frac{\sum_{s_1\in\smpOne}\old{p}_1(s_1)\old{m}_1\paren*{s_1,\old{\Delta}_1}}{\sum_{s_1\in\smpOne}\old{p}_1(s_1)}\\
\new{\Delta}_c &= \frac{\sum_{s_1\in\smpOne}\old{p}_c(s_1)\old{d}_c\paren*{s_1,\new{\mu}_c}}{\sum_{s_1\in\smpOne}\old{p}_c(s_1) \old{w}(s_1,\theta_c)}\\
\new{\Delta}_1 &= \frac{\sum_{s_1\in\smpOne}\old{p}_1(s_1)\old{d}_1\paren*{s_1,\new{\mu}_1}}{\sum_{s_1\in\smpOne}\old{p}_1(s_1) \old{w}(s_1,\theta_1)}\\
\new{\Gamma}_c &= \frac{\sum_{s_1\in\smpOne}\old{p}_c(s_1) \old{g}_c\paren*{s_1,\new{\mu}_c,\new{\Delta}_c}}{\sum_{s_1\in\smpOne}\old{p}_c(s_1)}
\end{align*}
\begin{align*}
\new{\Gamma}_1 &= \frac{\sum_{s_1\in\smpOne}\old{p}_1(s_1) \old{g}_1\paren*{s_1,\new{\mu}_1,\new{\Delta}_1}}{\sum_{s_1\in\smpOne}\old{p}_1(s_1)},
\end{align*}
where $\old{m}_{\ccdot}$, $\old{d}_{\ccdot}$, $\old{g}_{\ccdot}$ and $\old{w}_{\ccdot}$ ($\ccdot= c$ or $1$) are as defined in Table \ref{tab:algoQuantities}. Quantities $\old{p}_C$ and $\old{p}_1$ are defined as 
\begin{align}
    \old{p}_c(s_1)&=\frac{\old{\alpha}f_{\SN}(s_1;\old{\theta}_c)}{\old{\alpha}f_{\SN}(s_1;\old{\theta}_c)+ (1-\old{\alpha}) f_{\SN}(s_1;\old{\theta}_1)} \notag\\
     \old{p}_{1}(s_1)&=\frac{(1-\old{\alpha}) f_{\SN}(s_1;\old{\theta}_1)}{\old{\alpha}f_{\SN}(s_1;\old{\theta}_c)+ (1-\old{\alpha}) f_{\SN}(s_1;\old{\theta}_1)}. \label{eq:posterior}
\end{align}

\noindent The algorithm stops when the log-likelihood (Supplementary Materials) difference per data point falls under $10^{-8}$. False discovery rate at a threshold value $\tau$ is thereafter estimated as
\begin{align}
    \text{FDR}(\tau)
                 &=\frac{(1-\alpha) p(I_1>\tau)}{p(S_1>\tau)} \notag\\
                 &\overset{est}{=}\frac{(1-\alpha) \paren*{1-F_{\SN}(\tau;\theta_1)} }{\alpha \paren*{1-F_{\SN}(\tau;\theta_c)} + (1-\alpha) \paren*{1-F_{\SN}(\tau;\theta_1)}}. \label{eq:fdr}
\end{align}
To practically compute $F_{\SN}(\tau;\theta)$, we use an approximation of Owen's T function by \cite{young1974algorithm}.

\subsubsection{Parameter Initialization}
The initial parameters for the EM algorithm are estimated by partitioning the data and using the method of moments estimators for \SN\ distributions (Supplementary Materials). Precisely, $\smpOne$ is first partitioned into two sets separated by its median. The points below the median are then used to obtain a method of moments estimator of $\theta_1$ and the points above the median are used for $\theta_c$. Empirically, we observed that the signs of $\Delta_1$ and $\Delta_c$ do not change during the execution of the algorithm. To ensure that the entire parameter space is searched for an optimal fit, we run the algorithm four times covering all possible combinations of signs of $\Delta_1$ and $\Delta_c$, with the best fit chosen according to the value of the likelihood function. Parameter $\alpha$ is initialized at $0.5$.

\subsection{Top-two score skew normal mixture}
\label{sec:2s3dmix}
In the top-two score approach, referred to as 2SMix model, we model both first and second PSM score distributions as skew normal mixtures. Since the second score, $S_2$, can come from the correct, first incorrect or second incorrect match, we model its density as a three-component mixture. The complete model is specified as follows. 
\begin{align*}
     S_1 &\sim \alpha \SN(\theta_c) + (1-\alpha) \SN(\theta_1),\\
     S_2 &\sim \alpha \SN(\theta_1) + (1-\alpha-\beta) \SN(\theta_2) +  \beta \SN(\theta_c),
\end{align*}
\noindent 
where $\alpha,\beta \in [0,1]$ and $\alpha+\beta \leq 1$. The quintuple $\zeta=(\alpha, \beta, \theta_c, \theta_1,\theta_2)$ gives the parameters of the model. Observe that the two mixtures are tied via a shared parameter $\alpha$ because the fraction of the first incorrect PSMs in $\smpTwo$ must be identical to the fraction of correct PSMs in $\smpOne$. The fractions of correct PSMs in $\smpOne$ and $\smpTwo$ are further restricted by the fact that the total number of correct PSMs cannot exceed the sample size; i.e., $\alpha + \beta \leq 1$.

Unlike the top score only model, the parameters for the two score model cannot be obtained by using the existing skew normal mixture estimation methods because of parameter sharing between the two mixtures. We derive a novel EM algorithm for the maximum likelihood estimation of $\zeta$ from $\smpOne$ and $\smpTwo$.  Using $\new{\ }$ and $\old{\ }$ to accent the new and old parameter, respectively, the parameter update equations of the EM algorithm are as follows.
\begin{align*}
\new{\alpha} &= \frac{\sum_{s_1\in\smpOne} \old{p}_c(s_1) + \sum_{s_2\in\smpTwo} \old{r}_1(s_2)}{\abs*{\smpOne} + \abs*{\smpTwo}}\\
\new{\beta} &= \frac{\sum_{s_2\in\smpOne} \old{r}_c(s_2)}{\abs*{\smpTwo}}\\\new{\mu}_c &= \frac{\sum_{s_1\in\smpOne}\old{p}_c(s_1)\old{m}_c\paren*{s_1,\old{\Delta}_c} + \sum_{s_2\in\smpTwo}\old{r}_c(s_2)\old{m}_c\paren*{s_2,\old{\Delta}_c}}{\sum_{s_1\in\smpOne}\old{p}_c(s_1)+\sum_{s_2\in\smpTwo}\old{r}_c(s_2)}
\end{align*}
\begin{align*}
\new{\mu}_1 &= \frac{\sum_{s_1\in\smpOne}\old{p}_1(s_1)\old{m}_1\paren*{s_1,\old{\Delta}_1} + \sum_{s_2\in\smpTwo}\old{r}_1(s_2)\old{m}_1\paren*{s_2,\old{\Delta}_1}}{\sum_{s_1\in\smpOne}\old{p}_1(s_1)+\sum_{s_2\in\smpTwo}\old{r}_1(s_2)}\\
\new{\mu}_2 &= \frac{ \sum_{s_2\in\smpTwo}\old{r}_2(s_2)\old{m}_2\paren*{s_2,\old{\Delta}_2}}{\sum_{s_2\in\smpTwo}\old{r}_2(s_2)}\\
\new{\Delta}_c &= \frac{\sum_{s_1\in\smpOne}\old{p}_c(s_1)\old{d}_c\paren*{s_1,\new{\mu}_c} + \sum_{s_2\in\smpTwo}\old{r}_c(s_2)\old{d}_c\paren*{s_2,\new{\mu}_c}}{\sum_{s_1\in\smpOne}\old{p}_c(s_1)\old{w}(s_1,\theta_c)+\sum_{s_2\in\smpTwo}\old{r}_c(s_2)\old{w}(s_2,\theta_c)}\\
\new{\Delta}_1 &= \frac{\sum_{s_1\in\smpOne}\old{p}_1(s_1)\old{d}_1\paren*{s_1,\new{\mu}_1} + \sum_{s_2\in\smpTwo}\old{r}_1(s_2)\old{d}_1\paren*{s_2,\new{\mu}_1}}{\sum_{s_1\in\smpOne}\old{p}_1(s_1)\old{w}(s_1,\theta_1) + \sum_{s_2\in\smpTwo}\old{r}_1(s_2)\old{w}(s_2,\theta_1)}\\
\new{\Delta}_2 &= \frac{ \sum_{s_2\in\smpTwo}\old{r}_2(s_2)\old{d}_2\paren*{s_2,\new{\mu}_2}}{\sum_{s_2\in\smpTwo}\old{r}_2(s_2)\old{w}(s_2,\theta_2)}\\
\new{\Gamma}_c &= \frac{\displaystyle \sum_{s_1\in\smpOne}\old{p}_c(s_1)\old{g}_c\paren*{s_1,\new{\mu}_c,\new{\Delta}_c} + \displaystyle \sum_{s_2\in\smpTwo}\old{r}_c(s_2)\old{g}_c\paren*{s_2,\new{\mu}_c,\new{\Delta}_c}}{\sum_{s_1\in\smpOne}\old{p}_c(s_1)+\sum_{s_2\in\smpTwo}\old{r}_c(s_2)}\\
\new{\Gamma}_1 &= \frac{\displaystyle \sum_{s_1\in\smpOne}\old{p}_1(s_1)\old{g}_1\paren*{s_1,\new{\mu}_1,\new{\Delta}_1} + \displaystyle \sum_{s_2\in\smpTwo}\old{r}_1(s_2)\old{g}_1\paren*{s_2,\new{\mu}_1,\new{\Delta}_1}}{\sum_{s_1\in\smpOne}\old{p}_1(s_1)+\sum_{s_2\in\smpTwo}\old{r}_1(s_2)}\\
\new{\Gamma}_2 &= \frac{ \sum_{s_2\in\smpTwo}\old{r}_2(s_2)\old{g}_2\paren*{s_2,\new{\mu}_2,\new{\Delta}_2}}{\sum_{s_2\in\smpTwo}\old{r}_2(s_2)}
\end{align*}

\noindent where quantities $\old{m}_{\ccdot}, \old{d}_{\ccdot}, \old{g}_{\ccdot}$ and $\old{w}$ ($\ccdot= c, 1$ or $2$) are as defined in Table \ref{tab:algoQuantities}; 
$\old{p}_c, \old{p}_{1}$ are the same as those defined in Equation \ref{eq:posterior} and $\old{r}_c, \old{r}_{1}$ and $\old{r}_{2}$ are as defined below. 
\begin{align*}
    \old{r}_c(s_2) &=\frac{\old{\beta}f_{\SN}(s_2;\old{\theta}_c)}{\old{\alpha}f_{\SN}(s_2;\old{\theta}_1)+(1-\old{\alpha}-\old{\beta})f_{\SN}(s_2;\old{\theta}_2)+\old{\beta}f_{\SN}(s_2;\old{\theta}_c)}\\
     \old{r}_{1}(s_2) &=\frac{\old{\alpha}f_{\SN}(s_2;\old{\theta}_1)}{\old{\alpha}f_{\SN}(s_2;\old{\theta}_1)+(1-\old{\alpha}-\old{\beta})f_{\SN}(s_2;\old{\theta}_2)+\old{\beta}f_{\SN}(s_2;\old{\theta}_c)}\\
     \old{r}_{2}(s_2) &=\frac{(1-\old{\alpha}-\old{\beta})f_{\SN}(s_2;\old{\theta}_2)}{\old{\alpha}f_{\SN}(s_2;\old{\theta}_1)+(1-\old{\alpha}-\old{\beta})f_{\SN}(s_2;\old{\theta}_2)+\old{\beta}f_{\SN}(s_2;\old{\theta}_c)}.
 \end{align*}
As before, FDR is estimated according to Equation \ref{eq:fdr}.

\subsubsection{Parameter Initialization}
Similar to the parameter initialization for the top score mixture model, the top score is partitioned into two sets separated by its median and the points below the median are used to obtain a method of moments estimator of $\theta_1$ and the points above the median are used for $\theta_c$. The points of the second score corresponding to the top scores below the median, are used to obtain the initial estimate of $\theta_2$. To ensure that the entire parameter space is searched for an optimal fit, we run the algorithm eight times covering all possible combinations of signs of $\Delta_c$, $\Delta_1$ and $\Delta_2$, with the final fit selected based on the value of the likelihood function. Parameters $\alpha$ and $\beta$ are both initialized at $0.5$.

\begin{table}[ht]
\centering
\small
\caption{\small Datasets used for evaluation. MS-GF+ automatically sets the fragment ion tolerance based the chosen fragmentation method.}
\resizebox{\columnwidth}{!}
{\begin{tabular}{|c|l|l|l|p{1.5cm}|l|l|p{1.5cm}|}
\hline
 \thead{Dataset PXD} & Species                & Spectra & PSM   & Precursor Tolerance & Instrument           & Fragmentation Method & Missed Cleavages \\  \hhline{|=|=|=|=|=|=|=|=|}
PXD001179   & \emph{A.~thaliana}     & 116487  & 80894 & 10ppm     & LCQ/LTQ              & CID or by detection  & 1                \\
PXD006080   & \emph{D.~melanogaster} & 181749  & 72240 & 25ppm     & Orbitrap/FTICR/Lumos & CID or by detection  & 1                \\
PXD001481   & \emph{E.~coli}         & 59765   & 43217 & 10ppm     & LCQ/LTQ              & CID or by detection  & 1                \\
PXD012755   & \emph{H.~sapiens}      & 48754   & 48451 & 25ppm     & Orbitrap/FTICR/Lumos & CID or by detection  & 1                \\
PXD011988   & \emph{H.~sapiens}      & 35358   & 35176 & 25ppm     & Orbitrap/FTICR/Lumos & CID or by detection  & 1                \\
PXD013092   & \emph{M.~musculus}     & 86139   & 55312 & 15ppm     & Q-Exactive           & HCD                  & 2                \\
PXD001054   & \emph{M.~musculus}     & 69198   & 66113 & 15ppm     & Q-Exactive           & HCD                  & 2                \\
PXD001054   & \emph{M.~musculus}     & 57701   & 55312 & 15ppm     & Q-Exactive           & HCD                  & 2                \\
PXD001928   & \emph{S.~cerevisiae}   & 39284   & 38890 & 10ppm     & Q-Exactive           & CID or by detection  & 2                \\
PXD001928   & \emph{S.~cerevisiae}   & 37087   & 36402 & 10ppm     & Q-Exactive           & CID or by detection  & 2                \\ \hline

\multirowcell{4}{NIST \\Ion Trap} & \emph{C.~elegans}    & 67470  & 67308  & 25ppm & LCQ/LTQ & CID or by detection & 2 \\
                               & \emph{H.~sapiens}    & 340351 & 339857 & 25ppm & LCQ/LTQ & CID or by detection & 2 \\
                               & \emph{M. musculus}   & 149453 & 149325 & 25ppm & LCQ/LTQ & CID or by detection & 2 \\
                               & \emph{S.~cerevisiae} & 92608  & 92507  & 25ppm & LCQ/LTQ & CID or by detection & 2 \\ \hline
\end{tabular}
}
\label{tab:data}
\end{table}

\section{Experiments and Results}
\label{sec:results}
The experiments in this study were designed to investigate the properties and performance of the new methods. We first look at the accuracy of FDR estimation using the spectral libraries from NIST. We further use the libraries from NIST and datasets from PRIDE to evaluate the quality of the fit of the generative models and quantify the stability of FDR estimation.
Finally, we use an in-house experiment with diluted lysate of HeLa cells, with the total amount of digested protein ranging from 0.1ng to 100ng per analysis, to assess the robustness of FDR estimation to uncertainty and noise resulting from reduced levels of biological material and reduced levels of analytes.

\subsection{Datasets}
We used public and in-house data for model evaluation. The public data consisted of 4 ion trap datasets across four species from NIST spectral libraries \citep{stein1990national} and 10 datasets across six species from the PRIDE database \citep{Vizcaino2016}. All datasets are summarized in Table \ref{tab:data}. The protocols for generating in-house data and all relevant experimental details are described in Section \ref{sec:hela}.

\subsection{Database Search}
All searches were carried out using MS-GF+ \citep{Kim2014}, with search parameters identical to those from the publications associated with each dataset. Each dataset was searched against the corresponding species' proteomics database downloaded from UniProtKB \citep{Bairoch2005}. We carried out two searches. The first run was a target-decoy approach, where the decoy database was constructed by reversing tryptic peptides as proposed by \cite{Elias2007} and then concatenating these peptides to the target database. FDR at a score threshold $\tau$ was estimated as $\textrm{FDR}(\tau)=\frac{n_D(\tau)}{n_T(\tau)}$, where $n_D(\tau)$ is the number of top-scoring PSMs above $\tau$ that came from the decoy database and $n_T(\tau)$ is the number of top-scoring PSMs above $\tau$ that came from the target database. The second search was performed using the target database only and retaining up to 10 highest-scoring PSMs for each experimental spectrum. The results of these searches were used for decoy-free FDR estimation, as described in Section \ref{sec:methods}.

\begin{figure}[t]
\centering
    \resizebox{0.75\columnwidth}{!}{\includegraphics[width=.98\linewidth]{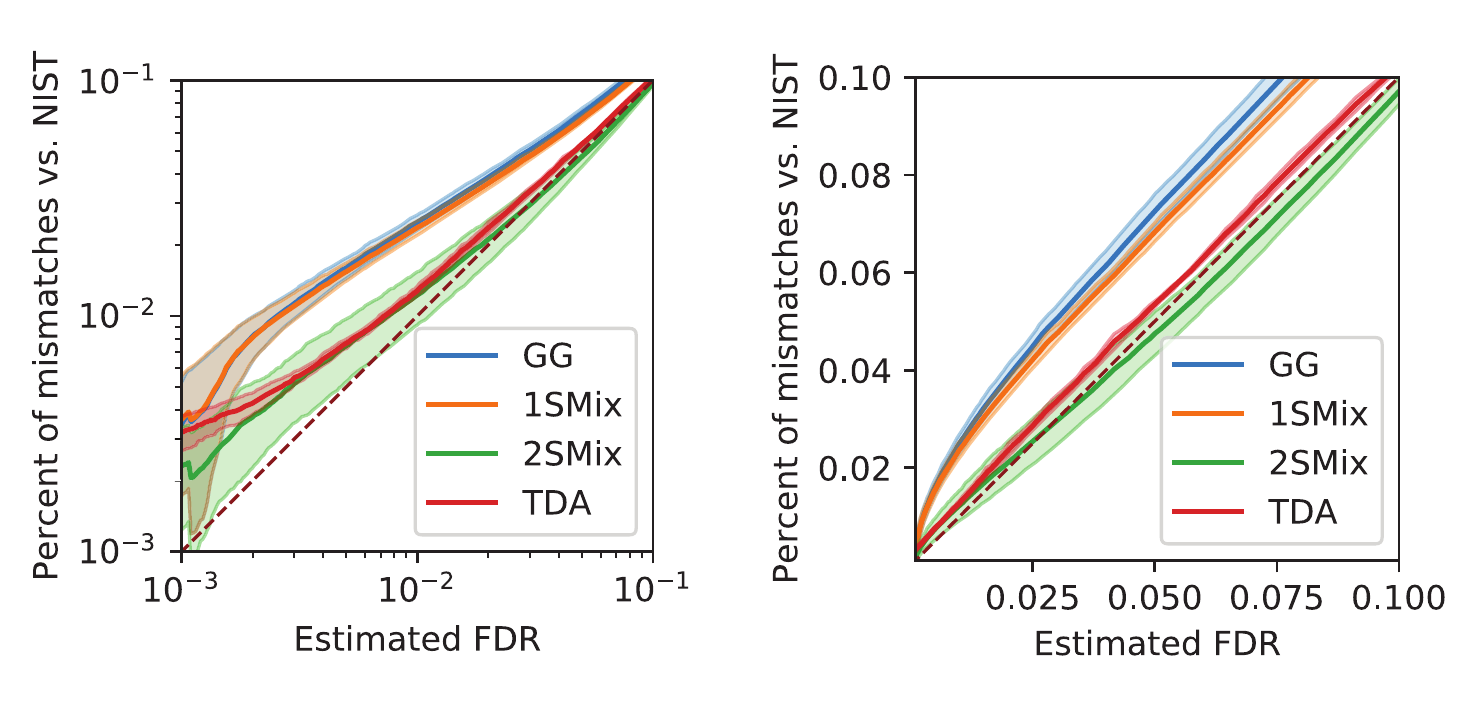}}
\caption{\small Fraction of mismatches in NIST library vs. estimated FDR. The closer to the identity line, the more accurate the estimation. Each curve is averaged over four NIST datasets, with the bands showing $68\%$ confidence intervals. On the left we show the log-scale to emphasize the range of more practical interest, while on the right we use linear scale.}
\label{fig:fdr}

\end{figure}

\subsection{Quality of FDR Estimates}
We searched NIST spectral libraries to establish the accuracy of FDR estimation. For each species and instrument platform, a NIST library consists of a set of consensus spectra, each associated with a peptide sequence, that can be considered as ground truth for our evaluation. After completing a search for which we estimated FDR, we computed the fraction of identified PSMs that did not match peptides from the NIST database as the true FDR and compared the two FDR values. This approach, however, has limitations. First, some peptides from NIST were not present in UniProtKB ensuring incorrect identifications in our searches whenever such a peptide received a sufficiently high score. Second, a peptide-spectrum pair in the NIST library may not always be a correct assignment in the first place because MS/MS searches may repeatedly lead to the same incorrect identifications due to database issues, peculiarities of the search parameters and software, or random chance. Third, we used the precursor mass tolerance of 25ppm that may be too stringent for the instrument types. This precursor tolerance was chosen to demonstrate the proof-of-principle of the developed approaches and show its potential applicability to data generated by different types of mass analyzers. Additionally, in some cases $k$ different peptides may be tied for the top score. We counted a $\frac{k-1}{k}$ fractional error in these cases if the correct peptide was among the $k$ peptides; otherwise, we counted a full error, regardless of the presence of the correct peptide in UniProtKB. An example of such a situation are peptides with leucine-to-isoleucine substitutions.

Figure \ref{fig:fdr} shows the estimated vs. true FDR averaged over four species from NIST in logarithmic and linear scale. We observe that the one-sample DFAs underestimate FDR, whereas the TDA and the two-sample DFA (2SMix) generates a curve closer to the diagonal line. Based on these results we conclude that the performance of TDA and the two-sample DFA is comparable, with the two-sample DFA having a slightly better performance in the low FDR range (0.001-0.01) and TDA having a slightly better performance in the high FDR range (0.01-0.1).

\begin{figure*}[ht!]
\small
\includegraphics[width=\linewidth]{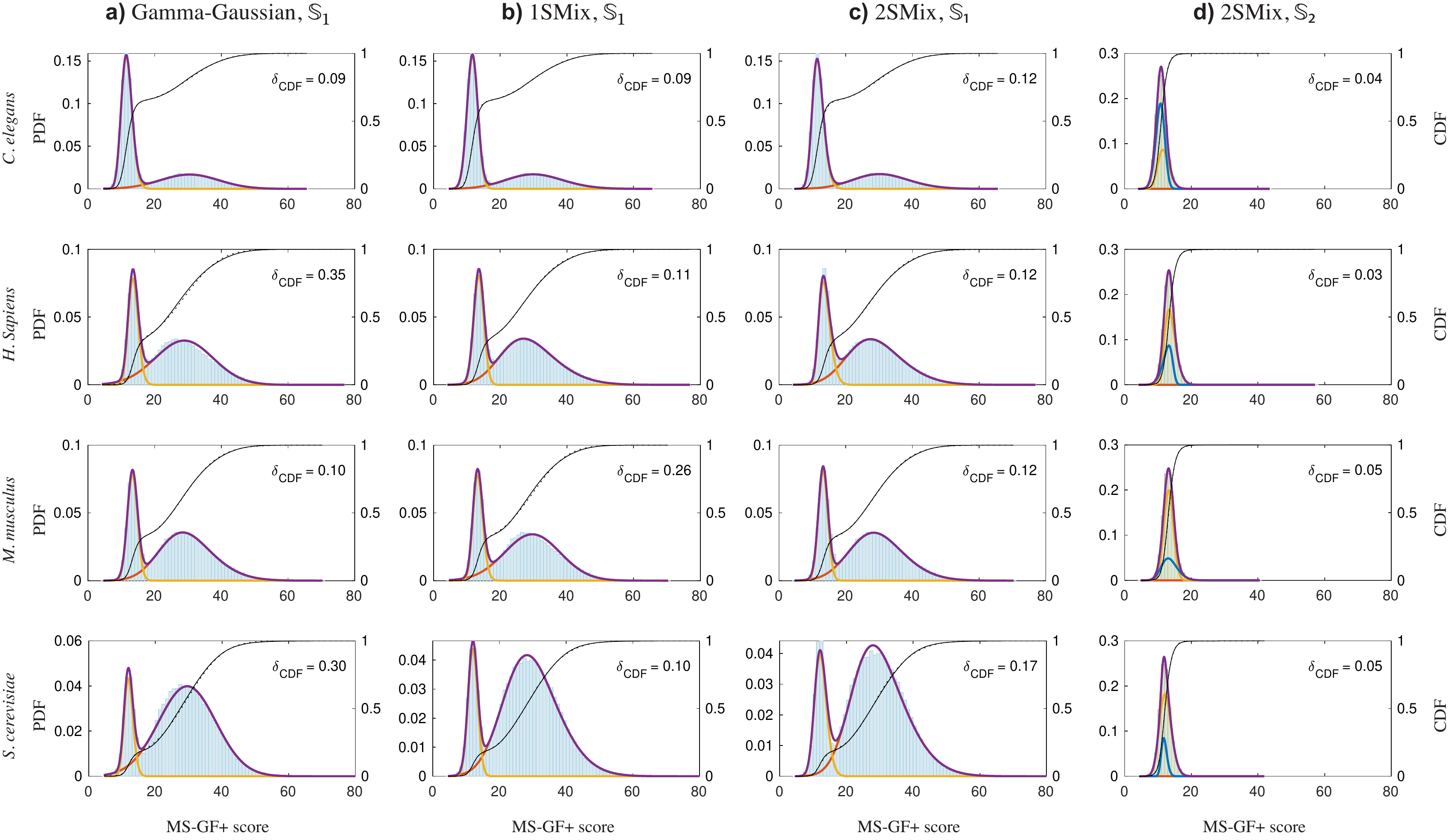}
\vspace{-2.5mm}
\caption{\small Model fitting on four NIST datasets. {\bf(a).} One-sample Gamma-Gaussian DFA estimation as proposed by \cite{Keller2002}, {\bf(b).} One-sample skew normal mixture 1SMix, {\bf(c,~d).} Two-sample skew normal mixture 2SMix. Histograms show score distributions $\smpOne$ (light blue) and $\smpTwo$ (light green), as a function of E-value. Purple densities superimpose estimated mixtures and their component distributions (yellow = top incorrect, blue = second-best incorrect, orange = correct). Estimated cdfs are shown in dotted black lines which that are mostly overlapping with the empirical cdfs shown in solid black lines.  Distances $\delta_{\textrm{CDF}}$, log-likelihoods and 1\% FDR thresholds are summarized in Table S1, Supplementary Materials.}
\label{fig:nist}
\end{figure*}

\begin{figure*}[hb!]
\vspace{-5mm}
\small
\includegraphics[width=\linewidth]{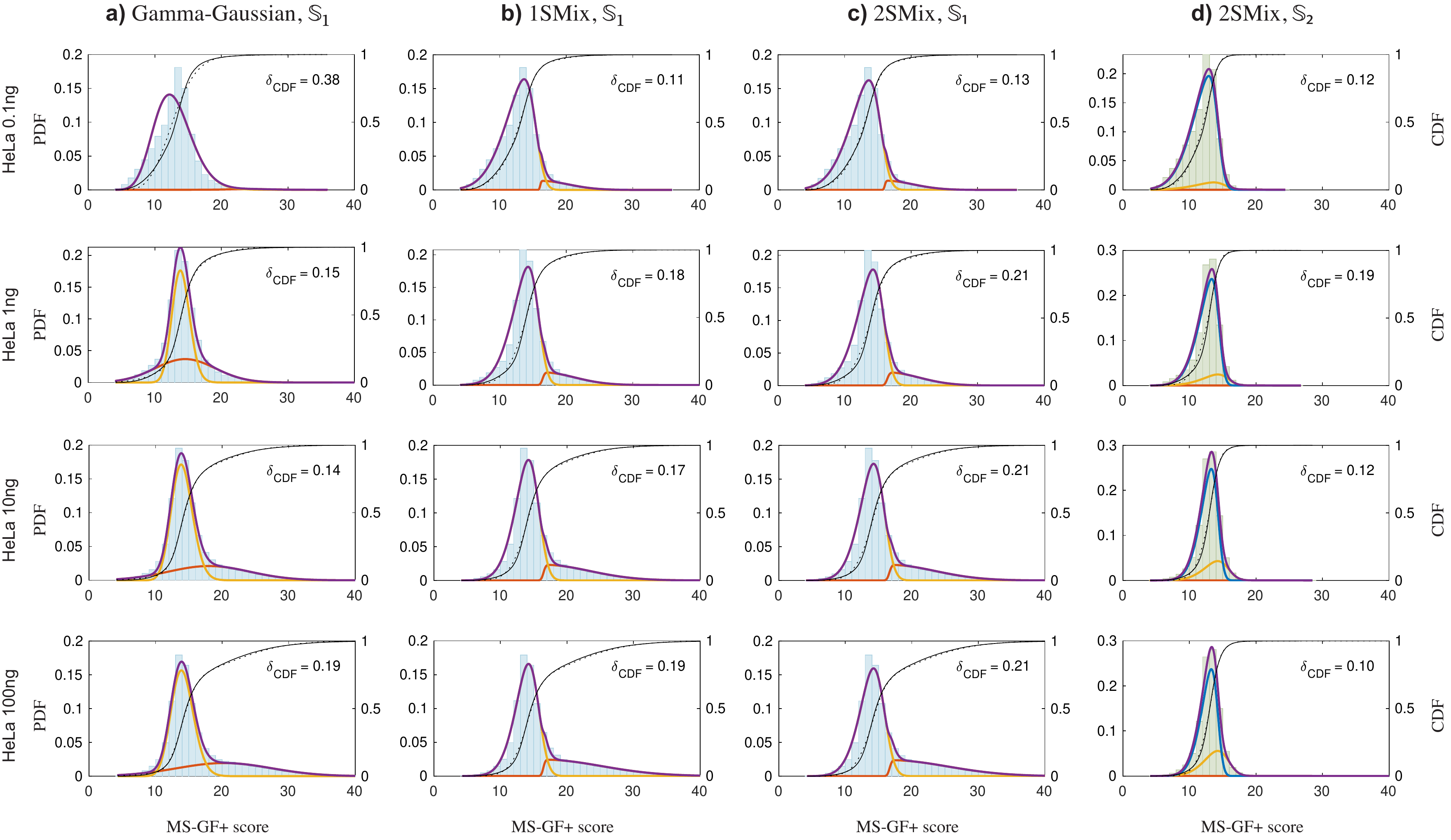}
\caption{\small Model fitting on four select HeLa cell datasets. {\bf(a).} One-sample Gamma-Gaussian DFA estimation as proposed by \cite{Keller2002}, {\bf(b).} One-sample skew normal mixture 1SMix, {\bf(c,~d).} Two-sample skew normal mixture 2SMix. Histograms show score distributions $\smpOne$ (light blue) and $\smpTwo$ (light green), as a function of E-value. Purple densities superimpose estimated mixtures and their component distributions (yellow = top incorrect, blue = second-best incorrect, orange = correct). Estimated cdfs are shown in dotted black lines which that are mostly overlapping with the empirical cdfs shown in solid black lines. Distances $\delta_{\textrm{CDF}}$, log-likelihoods and 1\% FDR thresholds are summarized in Table S1, Supplementary Materials.}
\label{fig:hela}
\end{figure*}

\subsection{Quality of the Fit}
Spectral libraries from NIST were also used to evaluate quality of the fit of the three DFAs. To do so, we plot the estimated probability density functions (pdfs) against the empirical score distributions in Figure \ref{fig:nist}. For each dataset, we evaluate the log-likelihood of the mixture sample $\smpOne$ and measure the cumulative distribution function (cdf) fit by computing $\delta_{\textrm{CDF}}$ as the unnormalized distance by \cite{Yang2019}, with $p=1$, between the empirical and estimated cdfs. For the two-sample DFA, we also evaluate the log-likelihood of the combined samples $\smpOne$ and $\smpTwo$ and additionally compute $\delta_{\textrm{CDF}}$ for $\smpTwo$. The distance between two cdfs was computed using the discrete cdf vectors of length $|\smpOne|$ or $|\smpTwo|$, as applicable. 

One-sample skew normal DFA improved the quality of the fit over Gamma-Gaussian DFA both in terms of log-likelihood and $\delta_{\textrm{CDF}}$ (Supplementary Materials). The log-likelihood values have been normalized by the sample size thus making the differences appear smaller than they are, whereas the $\delta_{\textrm{CDF}}$ measure appeared to be more in line with the visual inspection of the pdf fit. The two-sample skew normal DFA has somewhat reduced quality on $\smpOne$ compared to the one-sample skew normal DFA in both measures, but the high-quality fitting on $\smpTwo$ compensates for the difference. In addition, the quality of the fit of the second scores suggests that $\smpTwo$ indeed plays a role similar to that of the decoy database.

Datasets from PRIDE were additionally used to evaluate quality of the fit of the DFAs and to compare the cutoff values with TDA. The results of these experiments are summarized in Supplementary Materials for each of the 10 PRIDE datasets. Supplementary Table S1 gives summaries over these datasets. The findings on these datasets mirror those from NIST spectral libraries and increase confidence in strong performance of the two-sample DFA.

\subsection{Stability of FDR Estimates}
\label{sec:stability}
The stability of the FDR estimates was investigated using bootstrapping \citep{Efron1986}. In each of the $B=200$ bootstrap iterations, the spectra entering the search were sampled with replacement into an equal-sized set. After the database search, the 1\% FDR score threshold $\tau$ was estimated for each bootstrapped set using TDA and three DFAs. The variability in $\tau$ was then used to quantify stability of the estimates. 

\begin{figure}[h!]
\centering
    \resizebox{0.75\columnwidth}{!}{
\includegraphics[width=\linewidth]{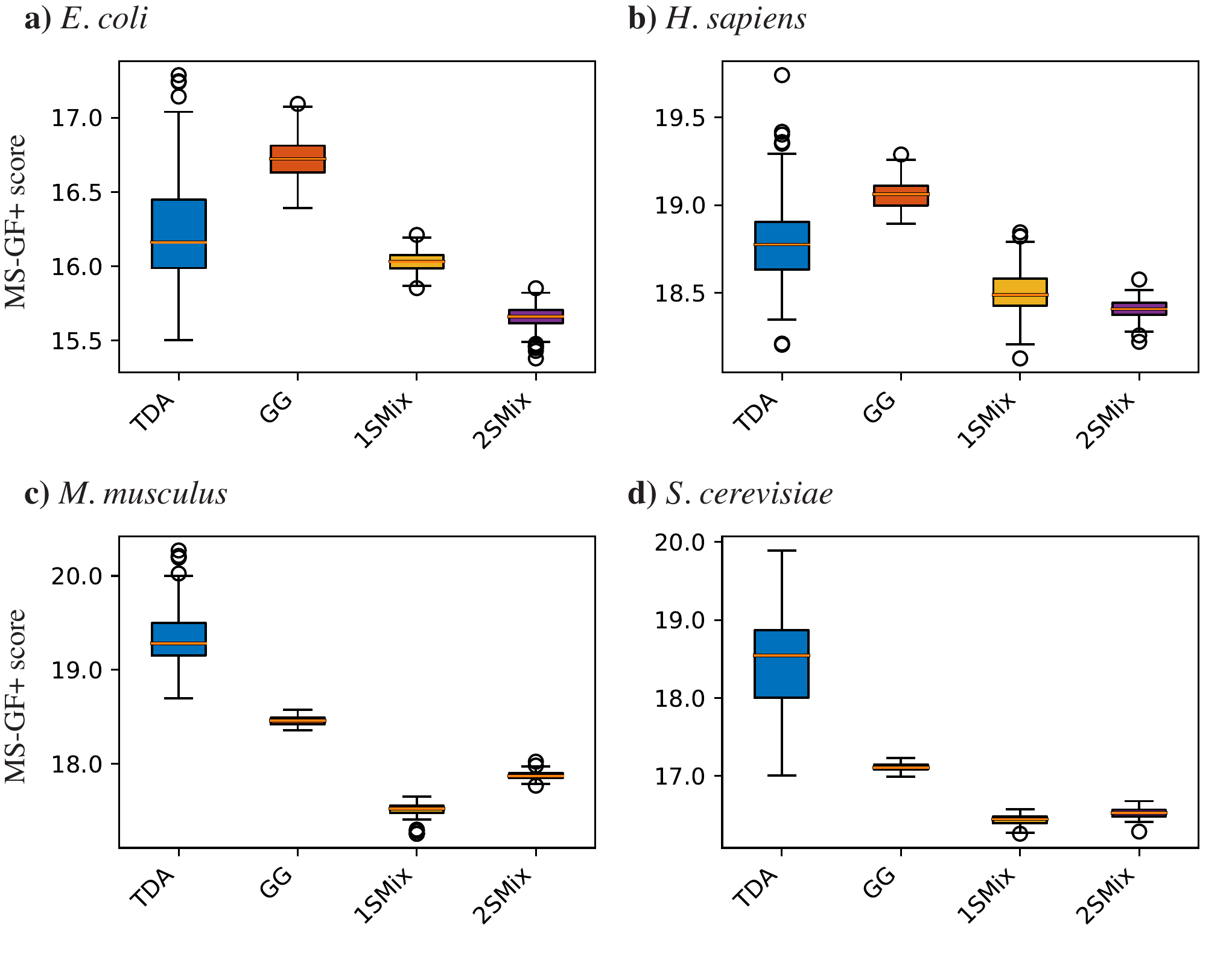}}
\caption{\small Stability of FDR estimates on four select datasets from PRIDE. The stability of estimates was evaluated using 200 bootstrapping iterations and measuring the 1\% FDR threshold in each of the iterations, as shown in the y-axis of each plot. The larger dispersion of established thresholds corresponds to lower stability of estimates.}
\label{fig:bootstrap}
\end{figure}

The stability of the four FDR estimation methods is compared in Figure \ref{fig:bootstrap} on four representative datasets from PRIDE. The results show that the TDA is generally less stable than any of the DFAs. This result is not entirely surprising given that the estimates of low FDR are often made based on a small number of decoy PSMs. Among DFAs, we find that one-sample DFAs were less stable than the two-sample DFA, suggesting that the two-sample DFA was able to capitalize on the existence of $\smpTwo$ to both improve and stabilize the estimate.

\begin{figure}[t]
\centering
    \resizebox{0.75\columnwidth}{!}{
\includegraphics[width=\linewidth]{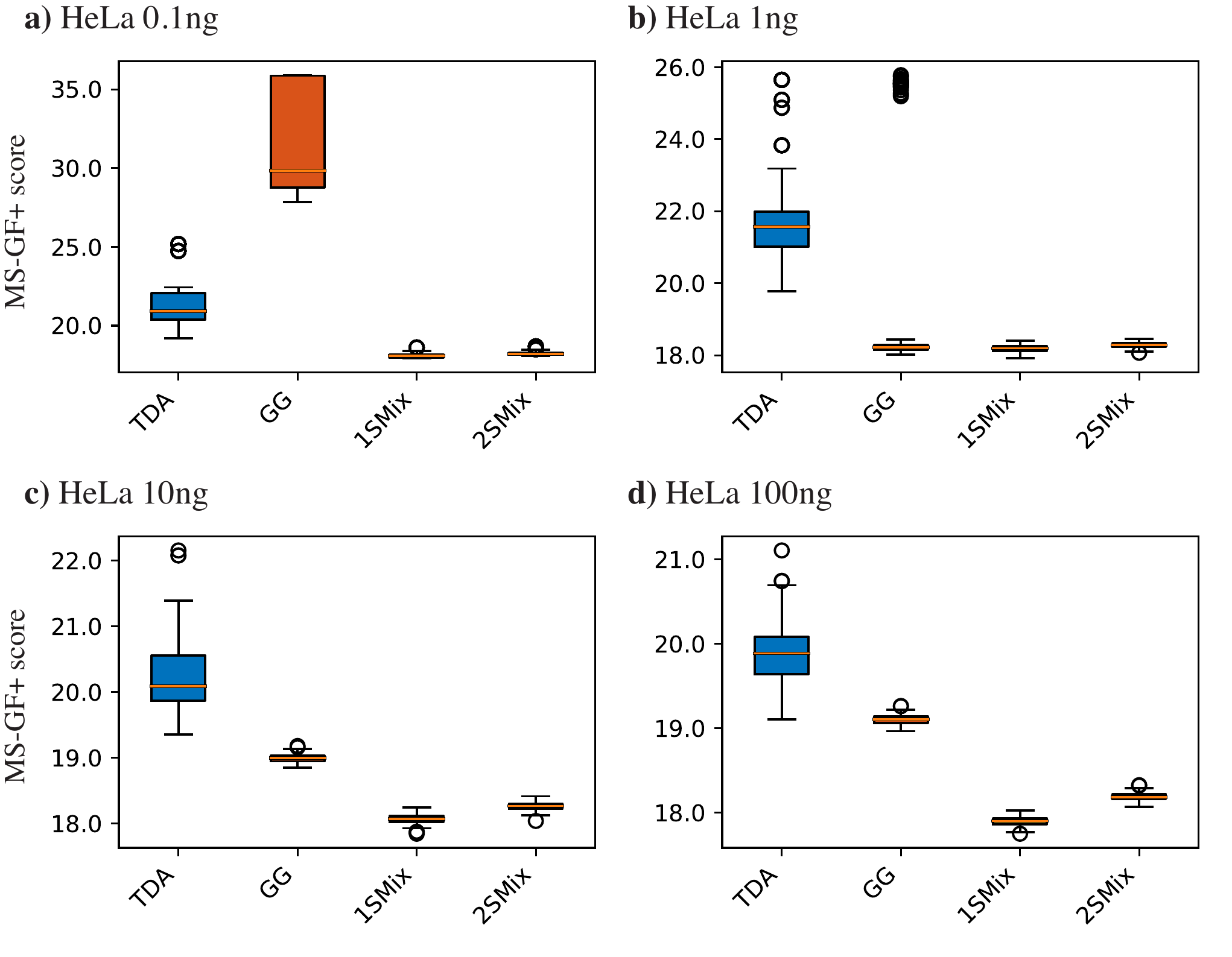}}
\caption{\small Stability of FDR estimates on four select datasets from the HeLa cell experiments. The stability of estimates was evaluated using 200 bootstrapping iterations and measuring the 1\% FDR threshold in each of the iterations, as shown in the y-axis of each plot. The larger dispersion of established thresholds corresponds to lower stability of estimates.}
\label{fig:bootstrap_hela}
\end{figure}

\subsection{HeLa Cell Digest Experiments}
\label{sec:hela}
\subsubsection{Experimental Setting}
To mimic the experiments requiring proteomics profiling of limited biomedical samples, we analyzed digested total lysate of cultured HeLa cells, which was selected as a representative high-complexity model sample. Sample aliquots were diluted to the desired concentration levels that corresponded to the total amount of digested protein ranging from 0.1ng to 100ng per analysis. The resulted specimens were analyzed using the conventional nano-flow liquid chromatography coupled with tandem mass spectrometry (nanoLC-MS/MS)-based approach, involving the separation conducted on a conventional 75$\mu$m inner diameter (ID) in-house bead-packed column. According to our estimates, the injected sample amounts corresponded to approximately 1--1,000 HeLa cells. The generated nanoLC-MS/MS data files were subjected to the analysis of spectral data, using the approach described next.

\subsubsection{LC-MS/MS Proteomics Analysis}
HeLa protein digest standard (P/N 88328, Thermo Fisher Scientific, Waltham, MA) was resuspended in 2\% formic acid to desired concentration levels. 0.1, 1, 10, 50 and 100ng of the HeLa digest aliquots were subjected to LC-MS/MS-based proteomics profiling. At least three technical replicates (i.e., replicate LC-MS/MS analyses of the same sample amount) were used across the whole study. The sample was loaded with the autosampler directly onto a self-packed column, which was made from a 75$\mu$m ID 360$\mu$m OD fused-silica capillary tubing (Molex, Polymicro Technologies, Phoenix, AZ) with a pulled tip filled with 20cm of 1.9$\mu$m ReproSil-Pur 120 C18-AQ (Dr. Maisch, Ammerbuch, Germany). Peptides were eluted at 150nL/min from the column using an UltiMate 3000 HPLC system (Thermo Fisher Scientific) with a 60 minute linear gradient from 1\% solvent B to 20\% solvent B (100\% acetonitrile, 0.1\% formic acid) mixed with solvent A (0.1\% formic acid in water). The eluent composition was changed from 20\% to 80\% of solvent B over 2 minutes and held constant for 3 minutes. Finally, the elution solvent composition was changed from 80\% solvent B to 99\% solvent A over 1 minute, and then held constant at 99\% of solvent A for 15 minutes. The application of a 2.3kV distal voltage electrosprayed the eluting peptides directly into an Orbitrap Fusion Lumos\texttrademark\ mass spectrometer equipped with a Nanospray Flex Ion Source (both Thermo Fisher Scientific). Mass spectrometer-scanning functions and HPLC gradients were controlled by the Xcalibur software (Thermo Fisher Scientific, v.4.1.50). The temperature of the ion transfer tube was set to 275\textdegree C. The mass spectrometer was set to scan MS1 at 120,000 resolution at $m/z$ 200 with an Automatic Gain Control (AGC) target set at 4e5 and for maximum injection time 50ms. The RF lens was set to 30\%. The scan range was $m/z$ 375-1500. Monoisotopic precursor selection mode was set to ``Peptide.'' For MS2, data-dependent acquisition mode was used. MS/MS spectra were acquired in the linear ion trap (rapid scan mode, HCD) with an AGC target of 3e4 and a maximum injection time (IT) at 35ms. The highest abundance peaks were analyzed by MS2 for a cycle time of 3 seconds and injecting ions using parallelization mode. Peptides were isolated with an isolation window of $m/z$ 1.6 and fragmented at higher-energy collisional dissociation energy of 28\%. Only ions with a charge state of 2 through 7 were considered for MS2. Dynamic exclusion was set at 30 seconds. The conversion of LC-MS .raw files to .mgf files was done using MSFileReader (v.2.2.62) and RawConverter v.1.1.0.23 \citep{He2015}. The default conditions for conversion were used, with one exception, charge states from 2 through 7 were used. The datasets were deposited in PRIDE (PXD020322).

\begin{figure}[]
\small
\centering
    \resizebox{0.75\columnwidth}{!}{
\includegraphics[width=\linewidth]{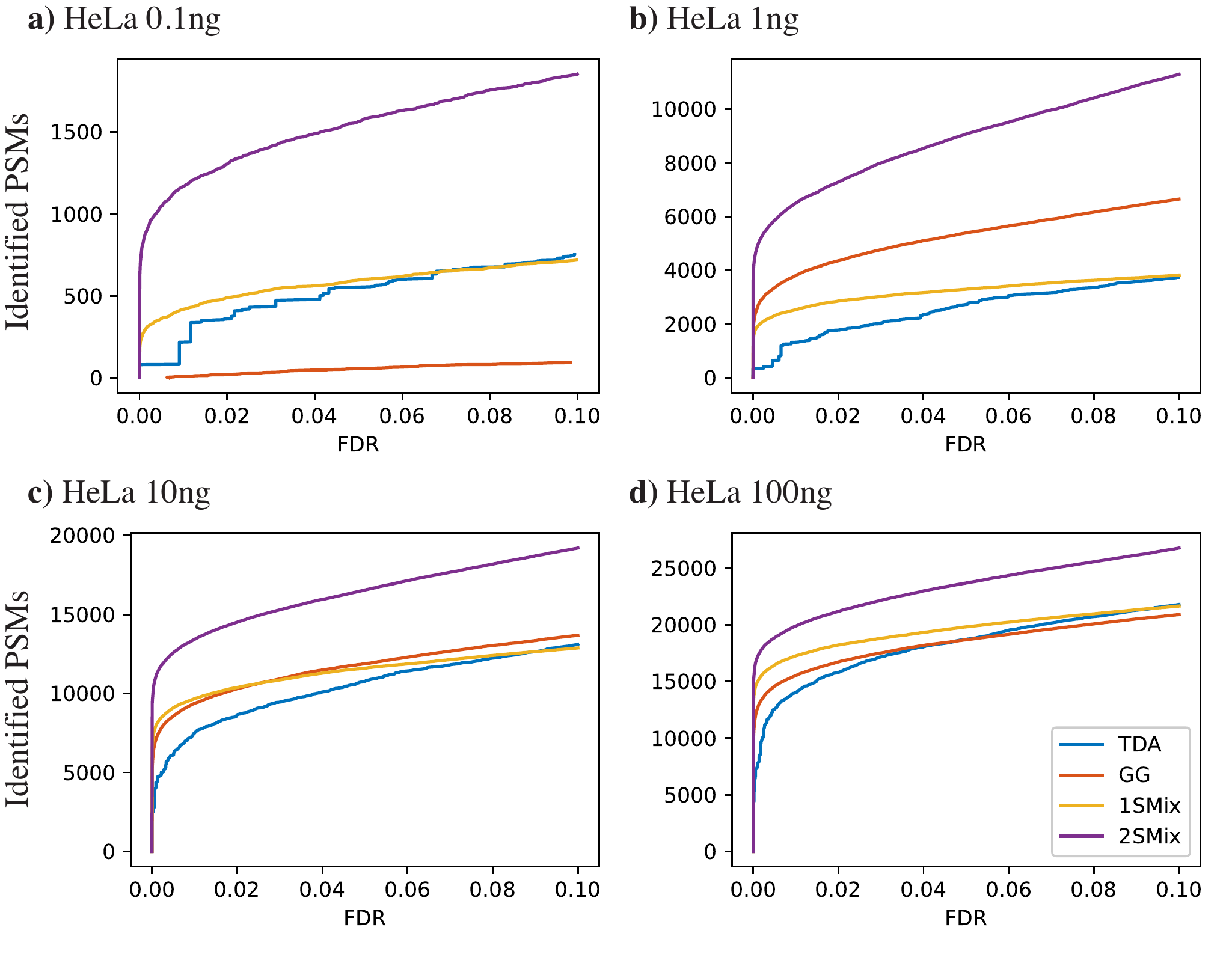}}
\vspace{-5mm}
\caption{\small The number of identified PSMs on the four select HeLa cell experiments at a specific FDR, separately estimated by each of the four individual methods.}
\label{fig:fdr_hela}
\end{figure}

\subsubsection{Results on HeLa Cell Experiments}
Figure \ref{fig:hela} shows a significantly improved fit of one- and two-sample skew normal mixtures compared to the Gamma-Gaussian mixture. Figure \ref{fig:bootstrap_hela} further visualizes stability of the 1\% FDR threshold in a bootstrapping experiment (as described in Section \ref{sec:stability}), suggesting that the two-sample skew normal mixture (2SMix) offers an attractive combination of fit and stability. Finally, Figure \ref{fig:fdr_hela} shows the number of identified PSMs as a function of estimated FDR in each of the experiments. It is worth noting here that the comparisons in Figure \ref{fig:fdr_hela} are not straightforward because each method estimates its own FDR and does so with different accuracy. However, we have previously demonstrated that TDA and the 2SMix DFA have comparable quality of FDR estimates (Figure \ref{fig:fdr}). In that light, we can more confidently infer an increased number of PSM identifications for the 2SMix DFA compared to TDA. Specifically, 687 more identifications for 0.1ng (+331\%), 2309 for 1ng (+168\%), 3488 for 10ng (+47\%), and 2469 for 100ng (+18\%) when averaged over the three replicates of each experiment.

Deep proteomic profiling of scarce biological and clinical samples is still a major challenge. The ability to qualitatively and quantitatively characterize thousands of proteins and their post-translational modifications present in limited samples (e.g., rare cell populations, microneedle biopsies, microsampled liquid biopsies, and even individually isolated single cells) is immensely important for getting new information in fundamental biology research and enabling novel diagnostic and prognostic studies \citep{Shao2018, Li2018, Lombard2019, Zhu2018, Li2015, Huffman2019}. However, the conventional nanoLC-MS/MS techniques fail to generate highly informative data at such sample levels. Since protein-derived analytes are at very low amounts in limited samples, the resulting MS and MS/MS spectra are generally sparse and low intensity. Interpretation of MS/MS fragmentation patterns resulting in correct peptide sequence identification and ultimately in in-depth protein and proteome characterization becomes a challenge using such low signal-to-noise-ratio and low fragmentation-efficiency spectra. Therefore, nanoLC-MS/MS analysis of limited samples typically results in a low conversion efficiency from tandem MS spectra to high-quality PSMs and a high FDR in peptide and protein identification, which in turn lead to limitations in quantitative analysis. We believe that the methodology proposed in this work improves the analysis of such samples.

\section{Conclusions}

Accurate FDR estimation has been one of the major computational challenges in bottom-up proteomics \citep{Nesvizhskii2010, Aggarwal2016} and is a key component of both peptide and protein identification \citep{Li2012a,Serang2012}. Although several approaches have been widely evaluated and used \citep{Keller2002, Elias2007, Kall2008, Jeong2012}, questions remain about their modeling assumptions, accuracy, stability, rigor and speed. The new types of experiments with low-amount analytes from limited samples, as the HeLa studies from our work, exemplify these challenges and require improved estimators. To address these challenges we proposed and evaluated new decoy-free methods for FDR estimation. Our methods rely on mixtures of skew normal distributions designed to model all component distributions. Importantly, our approaches eliminate the need to use a decoy database and, with it, the competition between peptides potentially present in the biological sample with those that are not. This is particularly evident in our two-sample DFA that relies on the score distribution of second-best PSMs associated with each spectrum and also models some level of dependence between first and second score distributions via parameter sharing and constraints.

The new mixture model methodology was extensively evaluated on public and in-house data. We show that one-sample DFAs are slightly inferior to TDA in terms of quality of FDR estimation, although they are faster and often more stable. On the other hand, our two-sample DFA offers an equivalent level of accuracy of FDR estimates as TDA, but with increased stability, improved speed, and slightly reduced cutoff thresholds that result in an increased number of PSM identifications (Section \ref{sec:results}). At the same time, the two-sample DFA retains methodological elegance of one-sample DFAs because skew normal distributions lend themselves to an efficient maximum likelihood optimization using expectation-maximization (Section \ref{sec:methods}). We believe that the new method will be applicable across a range of FDR estimation scenarios in bottom-up proteomics and beyond; e.g., with searches including post-translational modifications \citep{Fu2012}, cross-linked peptides \citep{Walzthoeni2012}, semi-tryptic peptides \citep{Alves2008}, \emph{de novo} searches \citep{Dancik1999, Frank2005}, small molecule searches \citep{Scheubert2017, Wang2018}.

\section*{Acknowledgements}
This work was supported by the National Institutes of Health awards 1R01GM103725 (PR), 1R01GM120272 (ARI) and 5R01CA218500 (ARI). We acknowledge Thermo Fisher Scientific for their support through a technology alliance. 

\bibliographystyle{plainnat}
\bibliography{refs-short}

\newpage
\setcounter{page}{1}

\appendix
\onecolumn

\end{document}